\documentclass[twocolumn,aps,superscriptaddress,nofootinbib]{revtex4-1}
\usepackage[dvips]{graphicx}
\usepackage{latexsym,amssymb,amsmath}
\usepackage{color}
\usepackage{enumerate}
\usepackage[bookmarksnumbered,bookmarksopen,colorlinks,citecolor=red,linkcolor=blue]{hyperref}
\usepackage{mathrsfs}
\usepackage{times}
\usepackage{soul}
\usepackage{csquotes}
\usepackage{multirow}
\usepackage{diagbox}
\usepackage{booktabs}
\usepackage[dvipsnames]{xcolor}

\graphicspath{{./figs/}}

\begin{document}

\title{Investigation of the tetraquark states $Qq\bar{Q} \bar{q}$ in the improved chromomagnetic interaction model}

\author{Tao Guo}
\affiliation{School of Mathematics and Physics, Chengdu University of Technology, Chengdu 610059, China}

\author{Jianing Li}
\affiliation{Department of Physics, Tsinghua University, Beijing 100084, China}

\author{Jiaxing Zhao}
\affiliation{SUBATECH, Universit\'e de Nantes, IMT Atlantique, IN2P3/CNRS, 4 rue Alfred Kastler, 44307 Nantes cedex 3, France}

\author{Lianyi He}
\affiliation{Department of Physics, Tsinghua University, Beijing 100084, China}

\date{\today}%

\begin{abstract}
In the framework of the improved chromomagnetic interaction model, we complete a systematic study of the $S$-wave tetraquark states $Qq\bar{Q}\bar{q}$ ($Q=c,b$, and $q=u,d,s$) with different quantum numbers, $J^{PC}=0^{+(+)}$, $1^{+(\pm)}$, and $2^{+(+)}$.
The mass spectra of tetraquark states are predicted and the possible decay channels are analyzed by considering both the angular momentum and $\mathcal{C}$-parity conservation.
The recently observed hidden-charm tetraquark states with strangeness, such as $Z_{cs}(3985)^-$, $X(3960)$, and $Z_{cs}(4220)^+$, can be well explained in our model. Besides, based on the wave function of each tetraquark state, we find that the low-lying states of each $Qq\bar{Q}\bar{q}$ configuration have a large overlap to the $Q\bar Q$ and $q\bar q$ meson basis, instead of $Q\bar q$ and $q\bar Q$ meson basis. This indicates one can search these tetraquark states in future experiments via the channel of $Q\bar Q$ and $q\bar q$ mesons.
\end{abstract}

\maketitle
%\tableofcontents

\section{Introduction}

Exotic hadrons, especially the heavy flavor exotic hadrons, provide a unique tool to study the nature of the strong force and the low-energy properties of Quantum Chromodynamics (QCD). Except for the open-charm tetraquark states, such as $X_{0,1}(2900)^0$~\cite{LHCb:2020bls} and $T_{cc}(3875)^+$~\cite{LHCb:2021vvq}, dozens of hidden-charm (-bottom) exotic states have been discovered since the observation of the hidden-charm state $X(3872)$ in 2003
by the Belle Collaboration~\cite{Belle:2003nnu}, see reviews~\cite{Chen:2016qju,Esposito:2016noz,Karliner:2017qhf,Guo:2017jvc,Brambilla:2019esw}. The most fascinating and unknown thing is the inner structure of these exotic hadrons. Theoretically, these exotic states can be mainly explained as multiquark states, which can be molecule states or compact tetraquark states~\cite{Esposito:2016noz,Wang:2013cya,Karliner:2017qhf}, hybrid states with $c\bar c$-$g$ configuration~\cite{Close:1994hc,Oncala:2017hop}, missing charmonium states, which their masses can be predicted by potential models but drastically modified by thresholds~\cite{Carames:2012th,Chen:2013wca,Swanson:2014tra}, a kinematical effect called triangle singularity~\cite{Nakamura:2019nwd,Guo:2019twa,Braaten:2019gfj}.
A clear probe to distinguish multiquark states from hybrid states or charmonia is the charged hidden-charm (-bottom) exotic state~\cite{Chiu:2006us}. 

In recent years, many hidden-charm (-bottom) exotic states with non-zero electric change such as $Z_c(3900)^+$~\cite{BESIII:2013ris,Belle:2013yex}, $Z_c(4025)^+$~\cite{BESIII:2013mhi}, $X(4100)^+$~\cite{LHCb:2018oeg}, $Z_c(4430)^+$~\cite{Belle:2007hrb,LHCb:2014zfx}, $Z_b(10610)^+$, and $Z_b(10652)^+$~\cite{Belle:2011aa} have been observed in experiments. From the decay, one can infer their quark constitutes are $[cu\bar c\bar d]$ or $[bu\bar b\bar d]$.
In the meantime, charged hidden-charm tetraquark states with strangeness are also been found experimentally.
In 2020, the BESIII Collaboration report a charged hidden-charm exotic structure with strangeness based on the processes of $e^+e^-\to K^+D_s^-D^{\ast 0}$ and $K^+D_s^{\ast-}D^{ 0}$~\cite{BESIII:2020qkh}.
Experimental analysis shows that the exotic state has a mass of $(3982.5_{-2.6}^{+1.8}\pm 2.1)$ MeV and a width of $(12.8_{-4.4}^{+5.3}\pm 3.0)$ MeV, which is close to the $D_s^-D^{\ast 0}$ and $D_s^{\ast-}D^{ 0}$ thresholds.
It is the first observed candidate of the charged hidden-charm tetraquark with strangeness, $[cs\bar{c}\bar{u}]$, named $Z_{cs}(3985)^-$. Next, the LHCb Collaboration observed an exotic state, $Z_{cs}(4000)^+$, with a mass of $(4003\pm6^{+4}_{-14})$ MeV and a width of $(131\pm15\pm16)$ MeV in the $J/\psi K^+$ invariant-mass spectrum~\cite{LHCb:2021uow}. Its quark composition is probably $[cu\bar{c}\bar{s}]$.
In addition, three new candidates named $Z_{cs}(4220)^+$, $X(4685)$, and $X(4630)$ are also observed with high significance in the $J/\psi K^+$ and $J/\psi\phi$ final states, respectively~\cite{LHCb:2021uow}.
Very recently, a near-threshold peaking structure referred to as $X(3960)$, is discovered by the LHCb Collaboration in the $D_s^+ D_s^-$ invariant mass spectrum~\cite{LHCb:2022vsv}. It's very likely a hidden-charm and hidden-strange tetraquark state, $[cs\bar{c}\bar{s}]$.
The best fit gives the mass and width of $X(3960)$ are $(3955 \pm 6 \pm 22)$ MeV and $(48\pm 17 \pm 10)$ MeV, respectively. The quantum number of this state is favored to be $I(J^{PC})=0(0^{++})$. In addition, a possible structure $X_0(4140)$ is observed in the $D_s^+D_s^-$ invariant mass spectrum~\cite{LHCb:2022vsv}.

On the theoretical side, the mass spectra of $Qq\bar Q\bar q$ states have been predicted by the potential model~\cite{Braaten:2014qka,Vijande:2007fc,Vijande:2007rf,Fernandez-Carames:2009qqw,Yang:2021zhe,Tiwari:2022azj}, QCD sum rule~\cite{Chen:2017dpy,Xin:2022bzt}, lattice QCD~\cite{Chiu:2006us,Bali:2011rd,Prelovsek:2013cra,Sadl:2021bme}, effective field theory~\cite{Hidalgo-Duque:2012rqv,Ke:2013gia,Cleven:2013sq,Dong:2021juy,Ge:2021sdq,Chen:2022dad}, and the chromomagnetic interaction (CMI) model~\cite{Wu:2018xdi}.
Because the CMI model~\cite{Silvestre-Brac:1992kaa,Buccella:2006fn,Hogaasen:2004pm,Cui:2005az,Maiani:2004uc,Stancu:2009ka,Guo:2011gu,Luo:2017eub,Wu:2017weo,Wu:2018xdi,Cheng:2020nho} only considers the short-range chromomagnetic interaction between constituent quarks and antiquarks, it is more suitable to describe the tight bound states. While for the heavy flavor exotic state, which contains more than one light quark (antiquark) probably has a large size, the chromoelectric contribution should be included. This comes to improved chromomagnetic interaction (ICMI) model~\cite{Hogaasen:2013nca,Weng:2018mmf,An:2021vwi,Guo:2021mja,Guo:2021yws,Weng:2021hje,Weng:2019ynv,Weng:2020jao,An:2022vtg,Weng:2022ohh,Liu:2022rzu}.
The ICMI model has been used to predict the mass spectra of open heavy-flavor tetraquark states~\cite{Guo:2021mja,Guo:2021yws,Weng:2021hje}, open and hidden heavy-flavor pentaquark states~\cite{Weng:2019ynv,Weng:2020jao,An:2022vtg}, and heavy-flavor dibaryons~\cite{Weng:2022ohh,Liu:2022rzu}. In this work, we investigated the masses, possible decay channels, and inner structures of changed and change-neutral, open- and hidden-strange tetraquark states $Qq\bar{Q}\bar{q}$ ($Q=c,b$, and $q=u,d,s$) via the ICMI model firstly.

The paper is organized as follows.
A brief introduction of the ICMI model and the wave functions of the tetraquark $Qq\bar{Q}\bar{q}$ systems in the color-spin space will be given in Sec.~\ref{sec2}.
In Sec.~\ref{sec3}, by substituting the effective masses and coupling strengths into the ICMI model, we calculated the mass spectra and wave functions of the $S$-wave tetraquark state $Qq\bar{Q}\bar{q}$.
In addition, the related analysis of possible decay channels and the inner structures are presented in this section.
We summarize in Sec.~\ref{sec.summary}.

%=========================
\section{Model description}
\label{sec2}
Analog to the meson and baryon, in this paper we study the heavy flavor tetraquark state $Qq\bar{Q}\bar{q}$. The state can be considered as the composition of a heavy quark $Q$, a heavy antiquark $\bar Q$, a light quark $q$, and a light antiquark $\bar q$ in the quark model, where $Q=c,b$, and $q=u,d,s$.
At the leading order, the strong interaction between constituent quarks (antiquarks) can be estimated by the one-gluon-exchange (OGE) potential. For $S$-wave tetraquark states, the spin-orbit angular momentum coupling part disappears. The total potential can be reduced to two parts~\cite{DeRujula:1975qlm,Oka:1989ud}
\begin{equation}
V_{ij}^{\text{OGE}}=V^{\text{cm}}_{ij} + V^{\text{ce}}_{ij},
\end{equation}
with the chromomagnetic interaction part,
\begin{equation}\label{cmp}
V^{\text{cm}}_{ij}=-{\alpha_s\pi \delta({\bf r}_{ij})\over 6m_im_j}{\lambda}_{i}^{c}\cdot{\lambda}_{j}^{c}\mbox{\boldmath{$\sigma$}}_i\cdot\mbox{\boldmath{$\sigma$}}_j,
\end{equation}
and the chromoelectric interaction part,
\begin{equation}\label{cep}
V^{\text{ce}}_{ij}={\alpha_s\over 4r_{ij}}{\lambda}_{i}^{c}\cdot{\lambda}_{j}^{c}.
\end{equation}
The parameter $m_i$ is the effective mass of the $i$th constituent quark, $\alpha_s$ is the running coupling constant, $r_{ij}=|{\bf r}_{ij}|=|{\bf r}_i-{\bf r}_j|$ is the spatial distance between the $i$-th and $j$-th quarks.
The notation ${\lambda}_{i}^{c}$ ($c=1, ..., 8$) are the Gell-Mann matrices acting on the color space of the $i$th quark, and the notation $\mbox{\boldmath{$\sigma$}}_i$ are the Pauli matrices on the spin space of the $i$th quark.
In addition, the $\lambda_i^c$ should be replaced by $-\lambda_i^{c*}$ if the subscript $i$ (or $j$) denotes an antiquark.
Integrating the spatial wave function part, we can get the ICMI model, which consists of the total mass of the constituent quarks, chromomagnetic interaction, and chromoelectric interaction.
Therefore, the effective Hamiltonian of a tetraquark system in the ICMI model reads~\cite{Hogaasen:2013nca}:
\begin{equation}\label{Ha0}
H=\sum\limits_{i=1}^4m_i+H_{\text{cm}} + H_{\text{ce}},
\end{equation}
where the chromomagnetic interaction term can be expressed as
 \begin{equation}\label{cm0}
H_{\text{cm}}=-\sum\limits_{i<j}v_{ij}{\lambda}_{i}^{c}\cdot{\lambda}_{j}^{c}
\mbox{\boldmath{$\sigma$}}_i\cdot\mbox{\boldmath{$\sigma$}}_j,
\end{equation}
and the chromoelectric interaction term can be expressed as
\begin{equation}\label{ce0}
H_{\text{ce}}=-\sum\limits_{i<j}c_{ij}{\lambda}_{i}^{c}\cdot{\lambda}_{j}^{c}.
\end{equation}
The model parameters $v_{ij}$ and $c_{ij}$ incorporate the effects of the effective mass of the constituent quarks, the spatial configuration of the tetraquark system, and the running coupling constant.
Considering the symmetry in the color-spin space, the chromoelectric interaction term and constituent quarks mass term can be consolidated into one term~\cite{Weng:2018mmf,Weng:2019ynv},
\begin{equation}\label{H0}
H_0\equiv \sum\limits_{i=1}^4m_i +H_{\text{ce}}=-\frac{3}{16}\sum\limits_{i<j}m_{ij}{\lambda}_{i}^{c}\cdot{\lambda}_{j}^{c},
 \end{equation}
 with an introduced parameter $m_{ij}=m_i+m_j+16c_{ij}/3$, which is related to the effective mass $m_i$ and $m_j$ of the constituent quarks and the coupling strength $c_{ij}$ of the chromoelectric interaction.
 Then, the Hamiltonian of the ICMI model can be simplified as,
 \begin{equation}
 H=H_0+H_{\text{cm}}.
 \label{Ha1}
 \end{equation}
The above parameters, such as $v_{ij}$ and $m_{ij}$ can be obtained by fitting the conventional hadron spectra.

Aiming to solve the eigen equations with given Hamiltonian~(\ref{Ha1}), we need to construct the wave function of $Qq\bar Q \bar q$ system, firstly.
For the tetraquark states $Q_1q_2 \bar{Q}_3\bar{q}_4$ where $Q=c,b$ and $q=u,d,s$, there are two types of decomposition of the wave function in color space based on the SU$(3)$ group theory. They physically correspond to two different configurations in color space, the diquark-antidiquark configuration labeled as $|(Q_1q_2)(\bar Q_3\bar{q}_4)\rangle$ and the meson-meson configuration labeled as $|(Q_1\bar{Q}_3)(q_2\bar{q}_4)\rangle$ (or $|(Q_1\bar{q}_4)(q_2\bar{Q}_3)\rangle$).
Taking into account the symmetry characteristics, these two configurations can be properly connected by a linear transformation.
It is convenient to see the total spin of the $S$-wave tetraquark states can be 0, 1, and 2, so the possible quantum numbers are $J^P=0^{+}$, $1^{+}$, and $2^{+}$.
Now, we can construct the color-spin wave function for given tetraquark states and we only show the results with the $|(Q_1\bar{Q}_3)(q_2\bar{q}_4)\rangle$ basis.

For the scalar tetraquark states with $J^P=0^{+}$, the color-spin basis wave functions $|\alpha_i\rangle$ ($i=1,2,3,4$) can be built as,
\begin{eqnarray}\label{1324basis0}
&&|\alpha_1\rangle\equiv |(Q_1\bar{Q}_3)^1_0\otimes(q_2\bar{q}_4)^1_0\rangle_0^1, \nonumber\\
&&|\alpha_2\rangle\equiv |(Q_1\bar{Q}_3)^1_1\otimes(q_2\bar{q}_4)^1_1\rangle_0^1, \nonumber\\
&&|\alpha_3\rangle\equiv |(Q_1\bar{Q}_3)^8_0\otimes(q_2\bar{q}_4)^8_0\rangle_0^1, \nonumber\\
&&|\alpha_4\rangle\equiv |(Q_1\bar{Q}_3)^8_1\otimes(q_2\bar{q}_4)^8_1\rangle_0^1,
\end{eqnarray}
where the superscripts and subscripts denote the total color and spin of $Q_1\bar{Q}_3$, $q_2\bar{q}_4$ subsystems, and the tetraquark $Q_1q_2 \bar{Q}_3\bar{q}_4$ systems, respectively.
We know the charge-neutral system has definite $\mathcal{C}$-parity. So, if $Q_1$ and $Q_3$, as well as $q_2$ and $q_4$ are of the same flavor. All the above color-spin bases have a positive charge conjugation, $J^{PC}=0^{++}$.

For the axial vector tetraquark states with quantum number $J^P=1^+$, the color-spin basis wave functions $|\beta_i\rangle$ ($i=1,2,...,6$) can be built as,
\begin{eqnarray}\label{1324basis1}
&&|\beta_1\rangle\equiv |(Q_1\bar{Q}_3)^1_0\otimes(q_2\bar{q}_4)^1_1\rangle_1^1, \nonumber\\
&&|\beta_2\rangle\equiv|(Q_1\bar{Q}_3)^1_1\otimes(q_2\bar{q}_4)^1_0\rangle_1^1, \nonumber\\
&&|\beta_3\rangle\equiv|(Q_1\bar{Q}_3)^1_1\otimes(q_2\bar{q}_4)^1_1\rangle_1^1, \nonumber\\
&&|\beta_4\rangle\equiv|(Q_1\bar{Q}_3)^8_0\otimes(q_2\bar{q}_4)^8_1\rangle_1^1, \nonumber\\
&&|\beta_5\rangle\equiv|(Q_1\bar{Q}_3)^8_1\otimes(q_2\bar{q}_4)^8_0\rangle_1^1, \nonumber\\
&&|\beta_6\rangle\equiv|(Q_1\bar{Q}_3)^8_1\otimes(q_2\bar{q}_4)^8_1\rangle_1^1.
\end{eqnarray}
Also, if $Q_1$ and $Q_3$, as well as $q_2$ and $q_4$ in the tetraquark $Q_1q_2 \bar{Q}_3\bar{q}_4$ systems are of the same flavor, the tetraquark state has a definite $\mathcal{C}$-parity.
The bases $|\beta_3\rangle$ and $|\beta_6\rangle$ do not change under the symmetry operation of charge conjugation, which gives $J^{PC}=1^{++}$, while $|\beta_1\rangle$, $|\beta_2\rangle$, $|\beta_4\rangle$, and $|\beta_5\rangle$ change the sign under the operation of charge conjugation, which gives $J^{PC}=1^{+-}$.

For $J^{P}=2^{+}$ states, the color-spin basis wave functions $|\gamma_i\rangle$ ($i=1,2$) are given by
\begin{eqnarray}\label{1324basis2}
&&|\gamma_1\rangle\equiv|(Q_1\bar{Q}_3)^1_1\otimes(q_2\bar{q}_4)^1_1\rangle_2^1, \nonumber\\
&&|\gamma_2\rangle\equiv|(Q_1\bar{Q}_3)^8_1\otimes(q_2\bar{q}_4)^8_1\rangle_2^1.
\end{eqnarray}
Similarly, each basis introduced above has a definite charge conjugation if $Q_1$ and $Q_3$, as well as $q_2$ and $q_4$ are of the same flavor. All the above color-spin bases of the tetraquark systems have a positive $\mathcal{C}$-parity.

The wave function $\Psi$ of the tetraquark state $Qq\bar{Q}\bar{q}$ with a given quantum number $J^{P}$ can be expressed as the superposition of the bases shown above,
\begin{equation}
\Psi=\sum_{i=1}^{N_{cs}} c_i|\kappa_i\rangle,
\end{equation}
where $N_{cs}$ represents the number of the color-spin basis and $c_i$ represents the amplitude for various color-spin basis, which satisfies the normalization condition $\sum_{i=1}^{N_{cs}} |c_i|^2=1$. Here, $|\kappa_i\rangle = |\alpha_i\rangle$, or $|\beta_i\rangle$, or $|\gamma_i\rangle$ depends on the quantum numbers.
With this wave function, we can obtain the matrix form of the Hamiltonian~(\ref{Ha1}), $\langle \Psi |H| \Psi \rangle$~\cite{Guo:2021mja}. The mass spectra of tetraquark states can be obtained by diagonalizing this matrix. And the probability $|c_i|^2$ can be used to analyze the possible decay channels of the tetraquark states.

%=============================
\section{Results and discussion}
\label{sec3}
\begin{figure*}[htbp]
\centering
\begin{minipage}{6.41cm}
    \includegraphics[width=0.98\textwidth,height=0.98\textwidth]{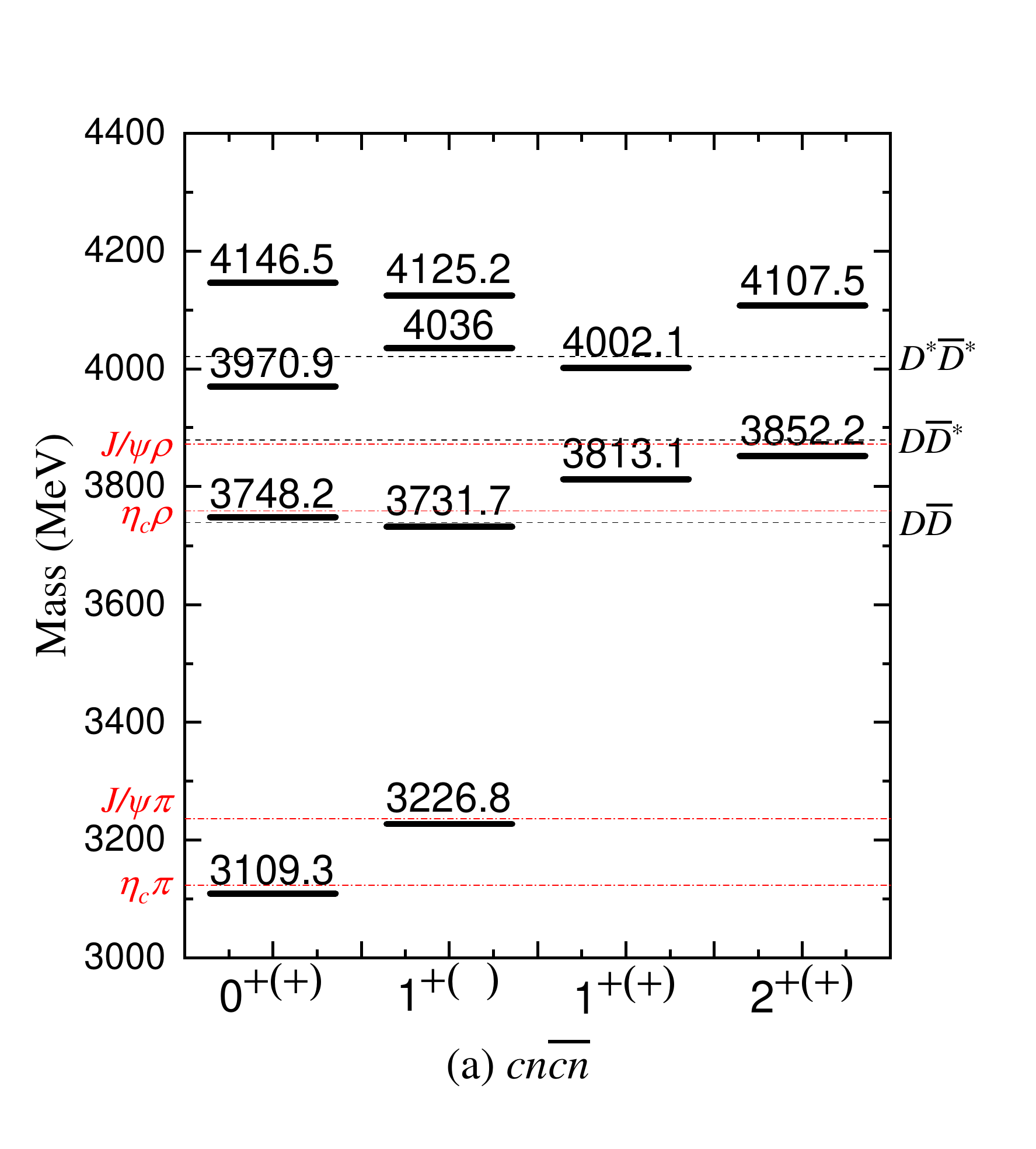}
\end{minipage}
\hspace{-0.85cm}
\begin{minipage}{6.41cm}
    \includegraphics[width=0.98\textwidth,height=0.98\textwidth]{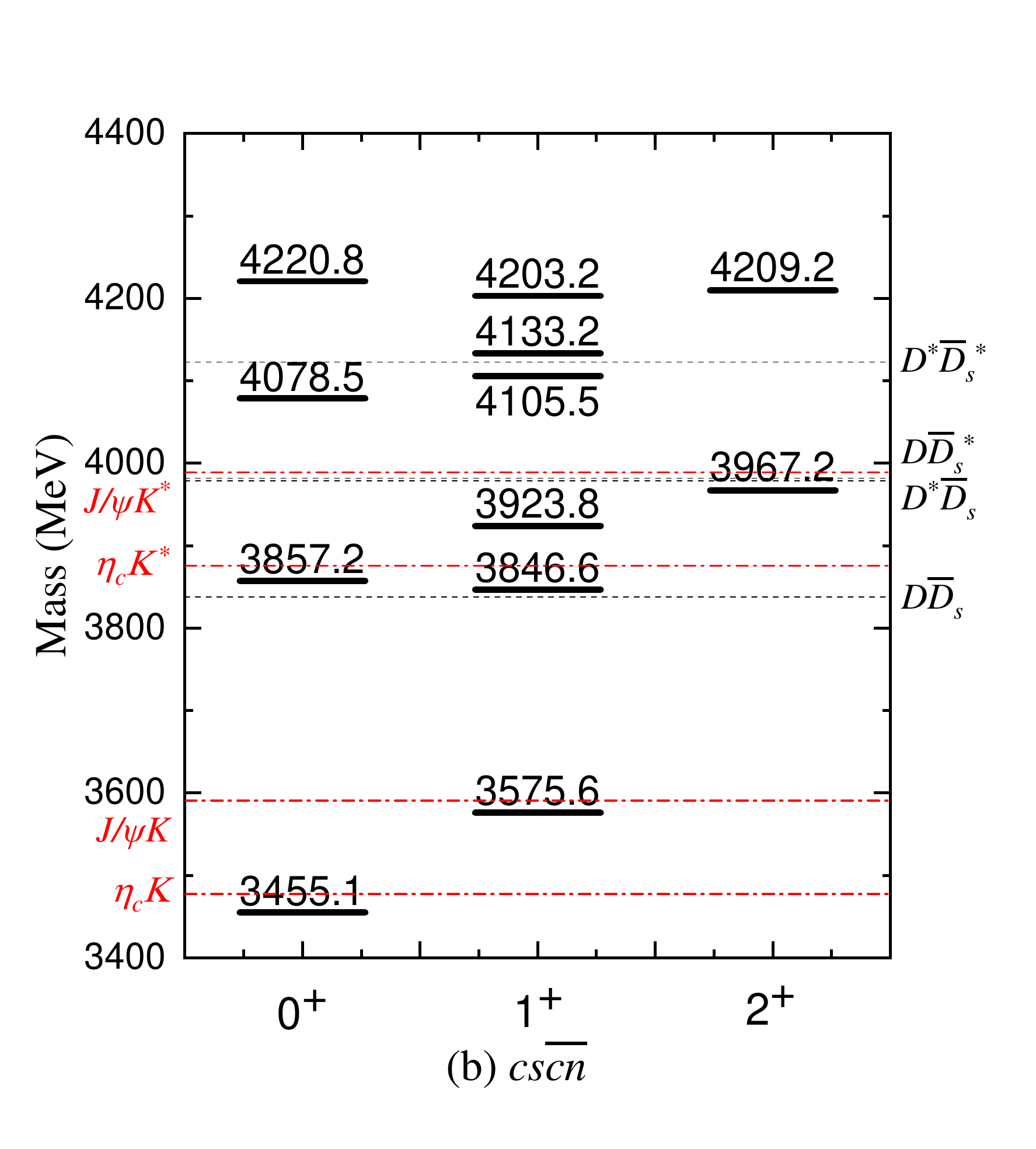}
\end{minipage}
\hspace{-0.85cm}
\begin{minipage}{6.41cm}
    \includegraphics[width=0.98\textwidth,height=0.98\textwidth]{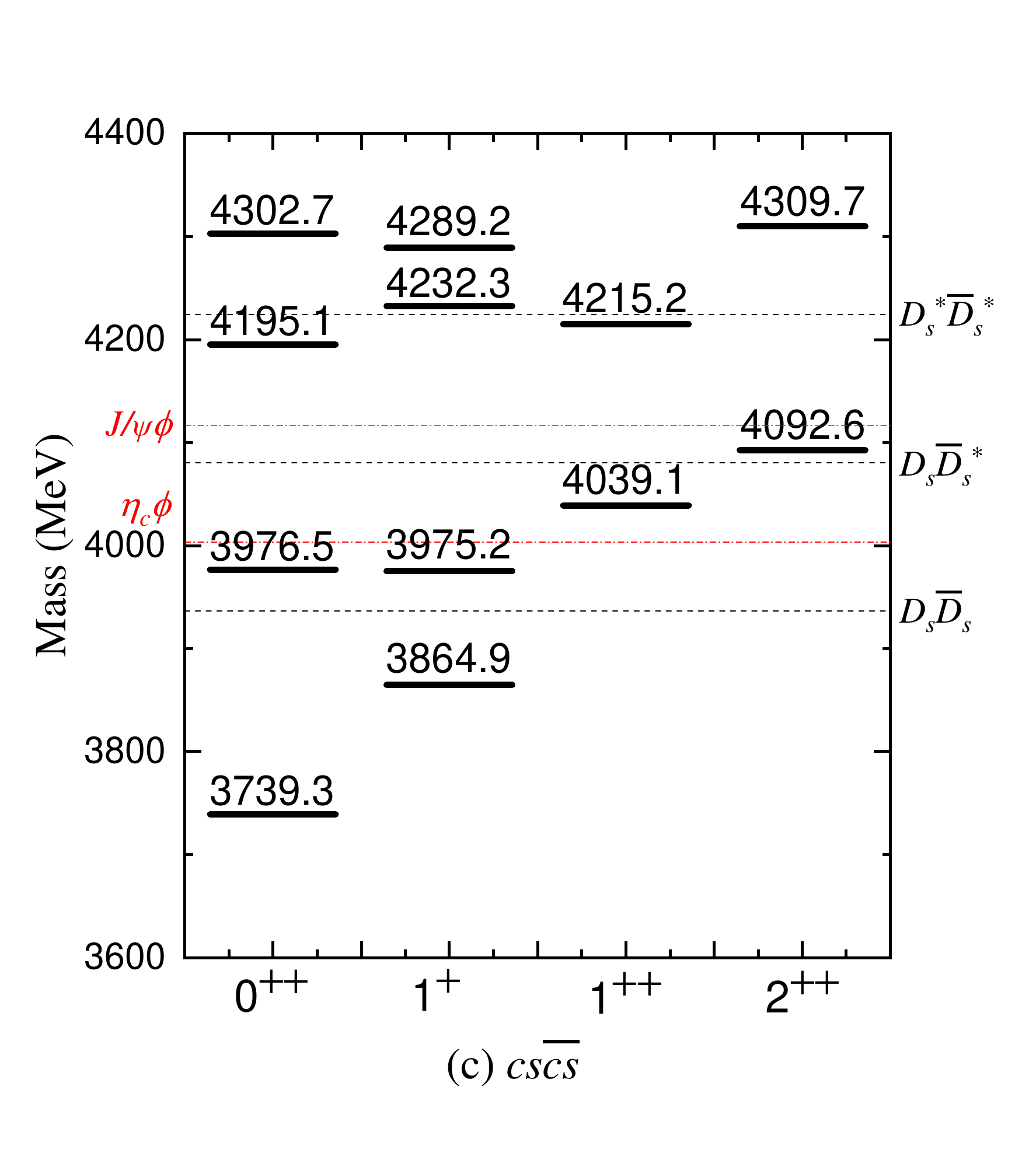}
\end{minipage}

\vspace{-0.5cm}
\caption{\label{fig1}
Mass spectra of $S$-wave tetraquark states (a) $cn\bar{c}\bar{n}$ ($n=u,d$), (b) $cs\bar{c}\bar{n}$, and (c) $cs\bar{c}\bar{s}$ with different quantum numbers.
The black dashed and red dot-dashed lines demonstrate all possible meson-meson thresholds.}
\end{figure*}

\begin{figure*}[htbp]
\centering
\begin{minipage}{6.41cm}
    \includegraphics[width=0.98\textwidth,height=0.98\textwidth]{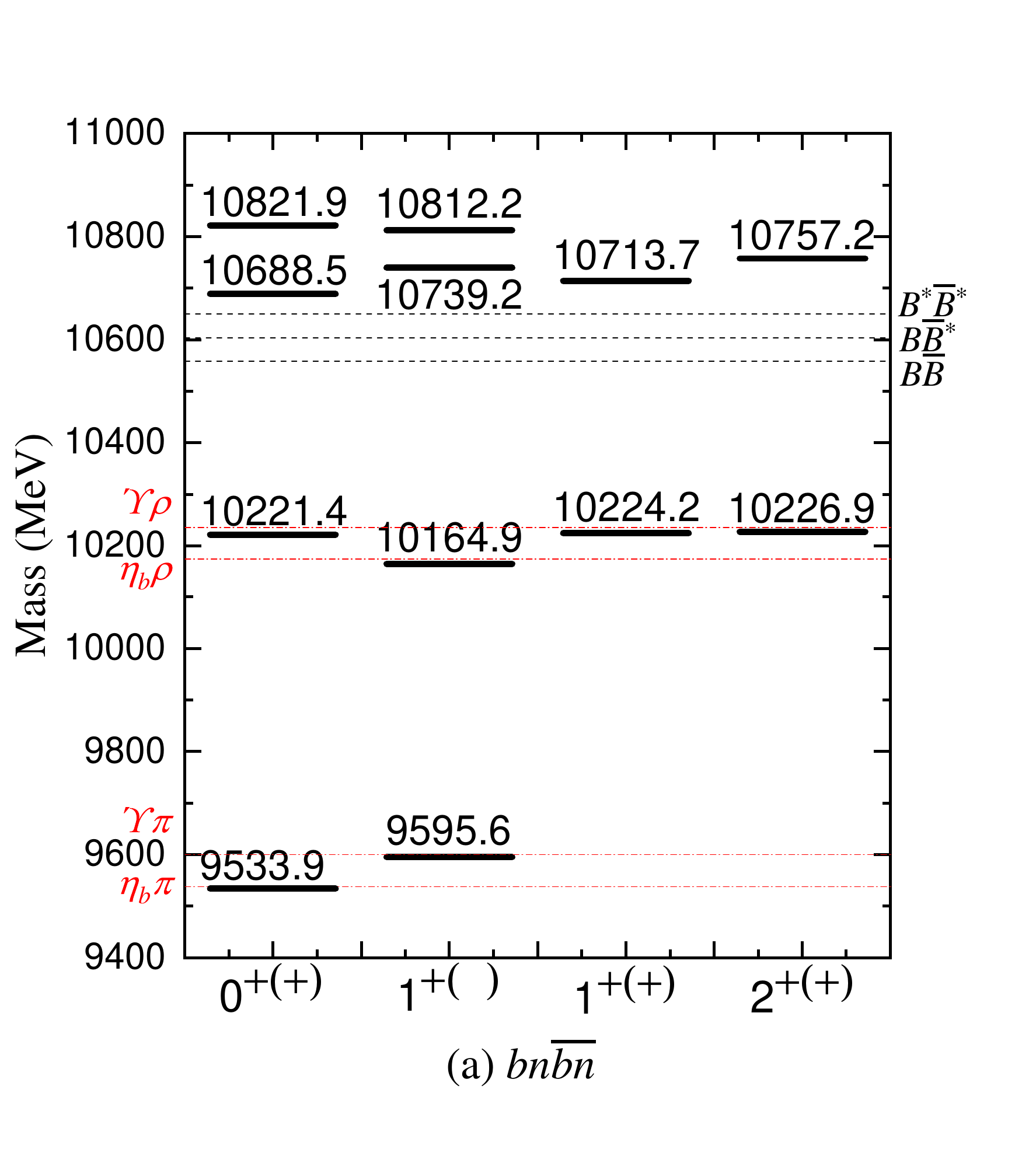}
\end{minipage}
\hspace{-0.85cm}
\begin{minipage}{6.41cm}
    \includegraphics[width=0.98\textwidth,height=0.98\textwidth]{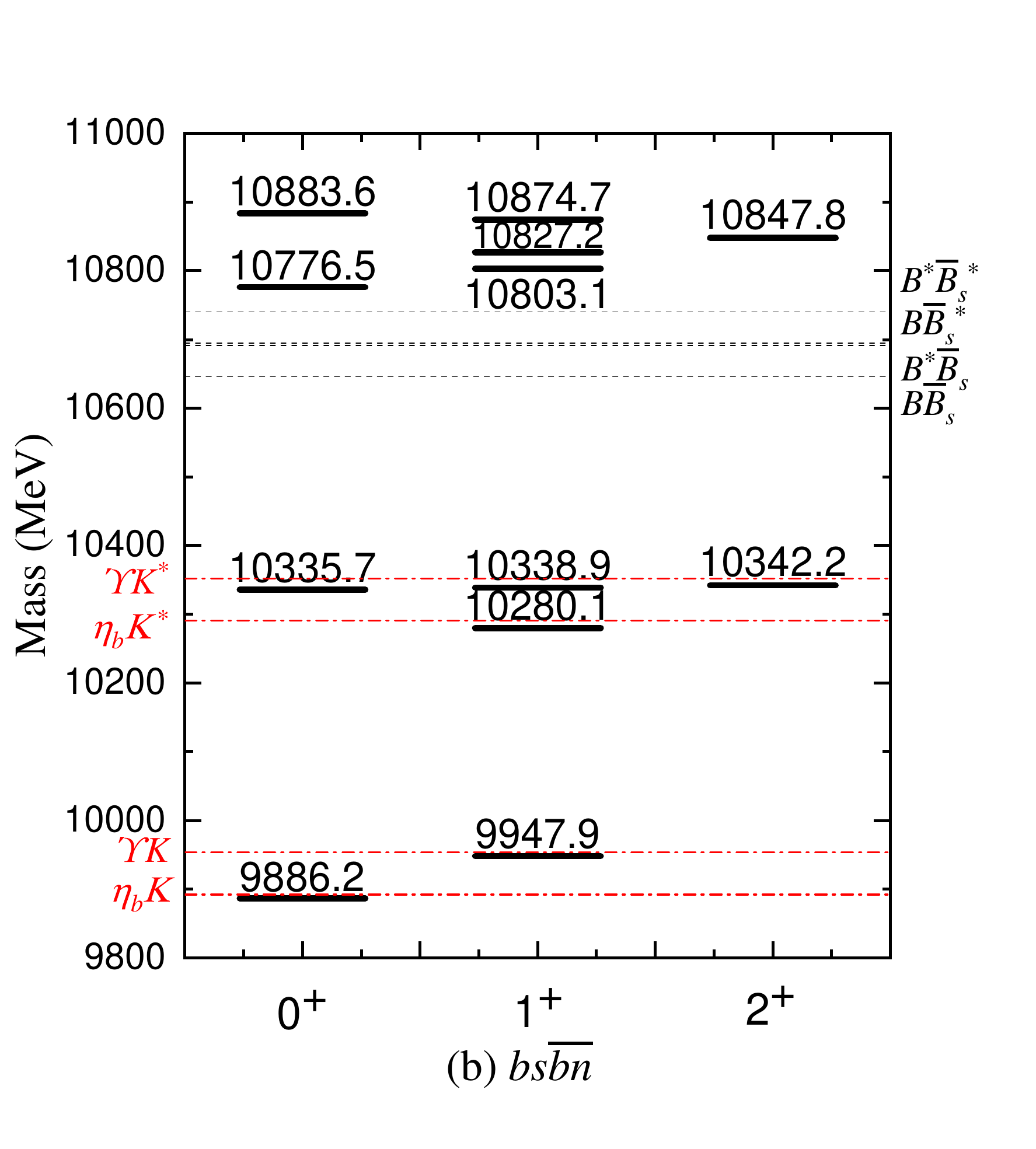}
\end{minipage}
\hspace{-0.85cm}
\begin{minipage}{6.41cm}
    \includegraphics[width=0.98\textwidth,height=0.98\textwidth]{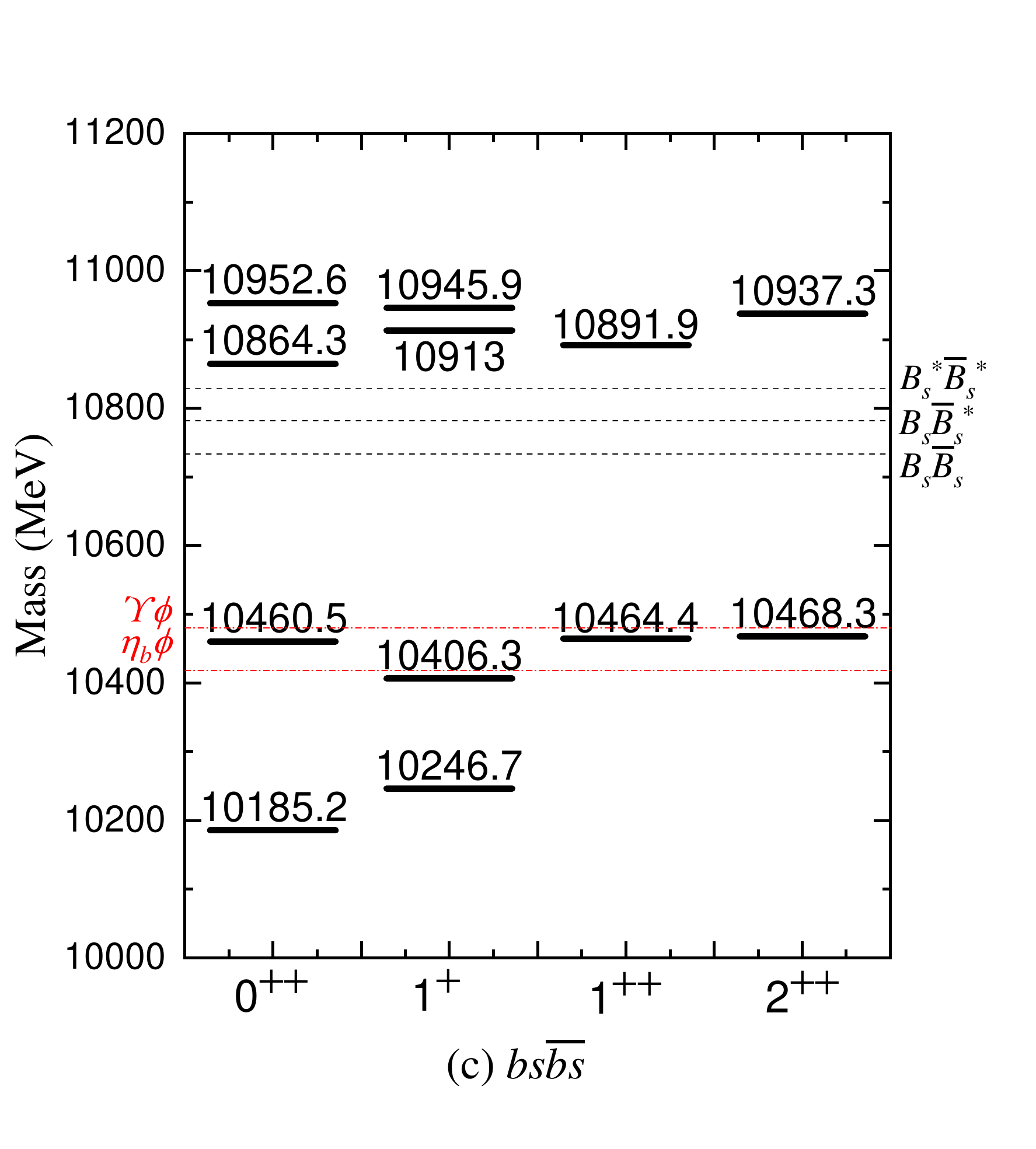}
\end{minipage}

\vspace{-0.5cm}
\caption{\label{fig2}
Mass spectra of $S$-wave tetraquark states (a) $bn\bar{b}\bar{n}$ ($n=u,d$), (b) $bs\bar{b}\bar{n}$, and (c) $bs\bar{b}\bar{s}$ with different quantum numbers. The black dashed and red dot-dashed lines demonstrate all possible meson-meson thresholds.}
\end{figure*}

\begin{figure*}[htbp]
\centering
\begin{minipage}{6.41cm}
    \includegraphics[width=0.98\textwidth,height=0.98\textwidth]{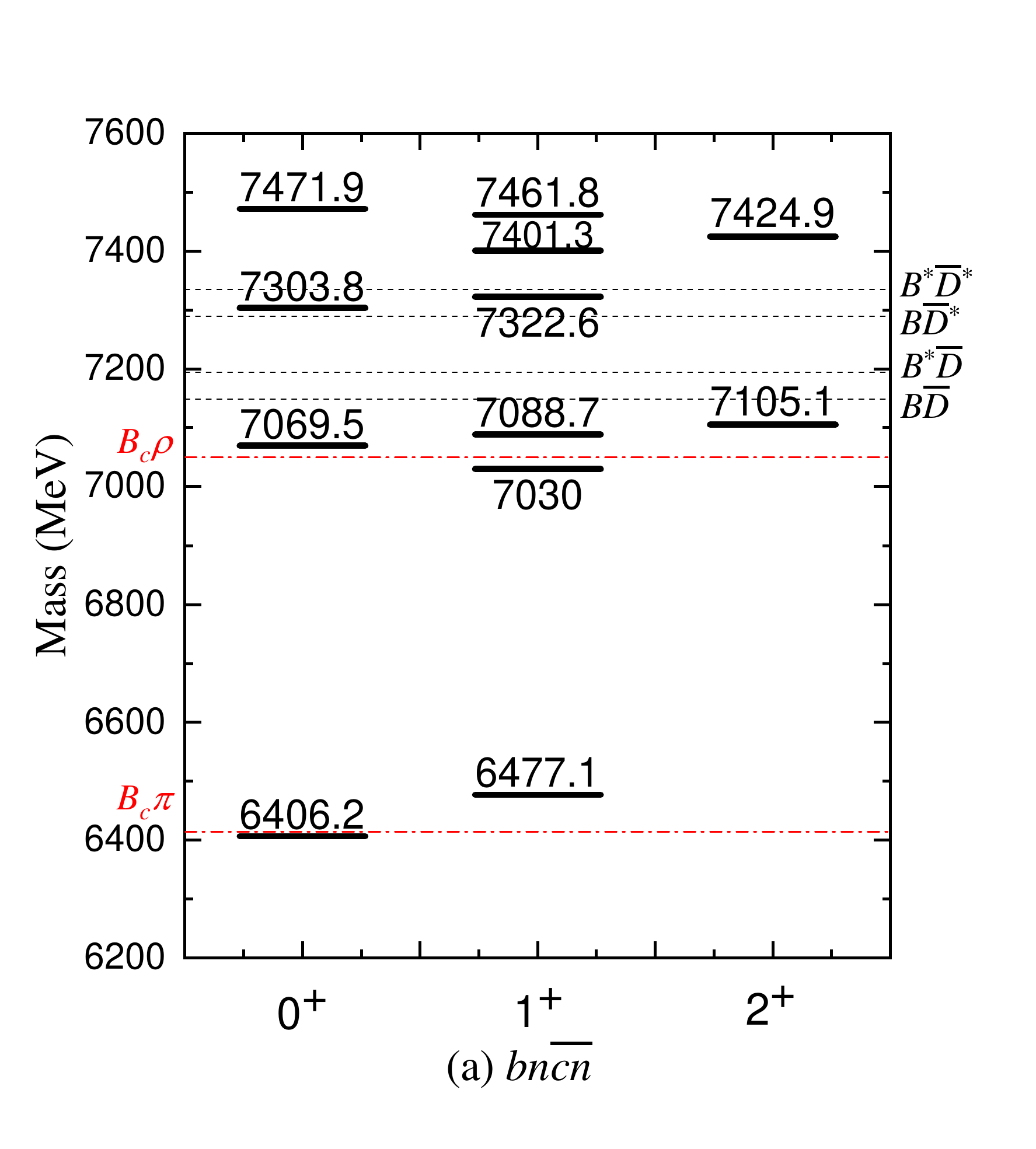}
\end{minipage}
\hspace{-0.85cm}
\begin{minipage}{6.41cm}
    \includegraphics[width=0.98\textwidth,height=0.98\textwidth]{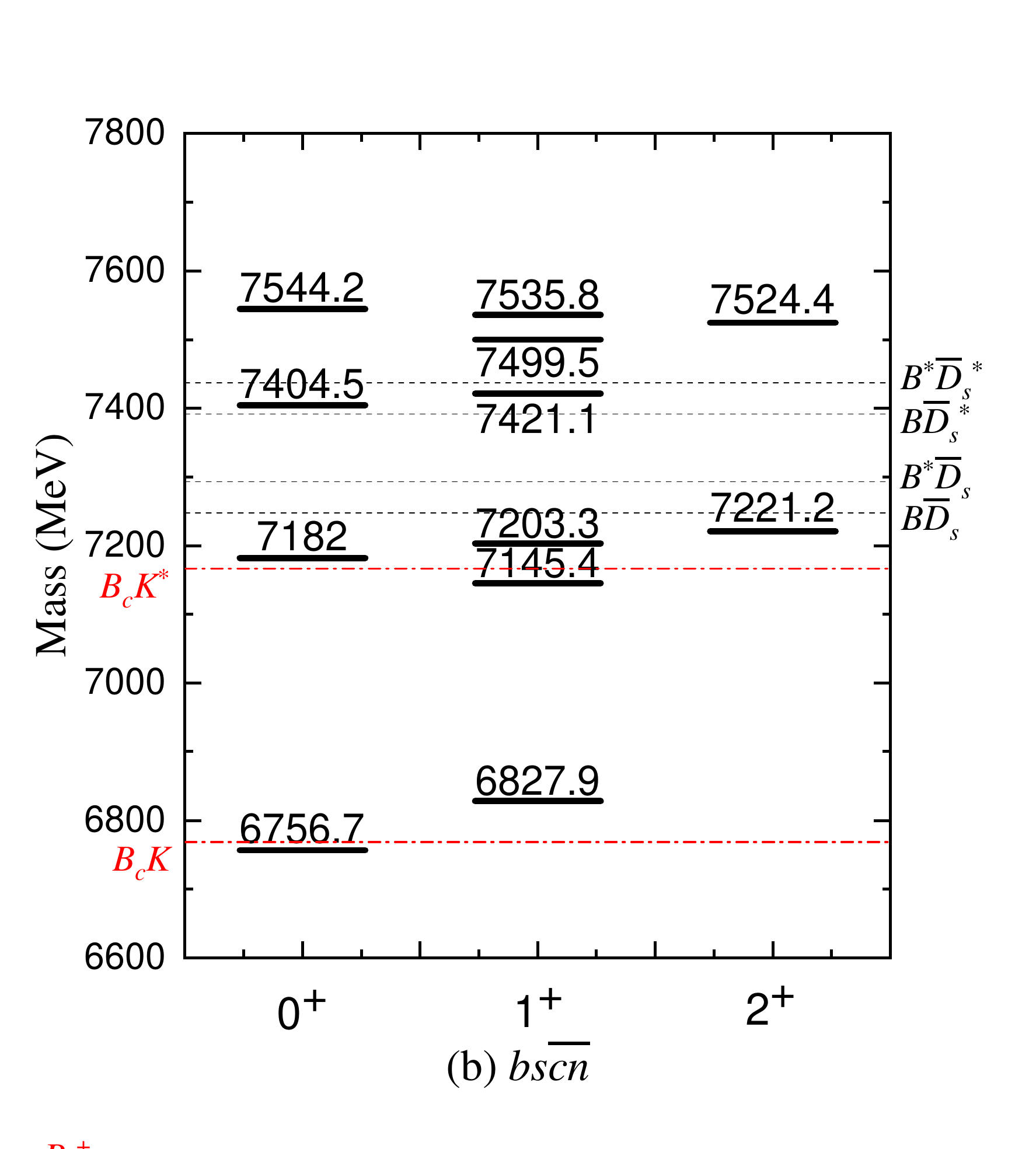}
\end{minipage}
\hspace{-0.85cm}
\begin{minipage}{6.41cm}
    \includegraphics[width=0.98\textwidth,height=0.98\textwidth]{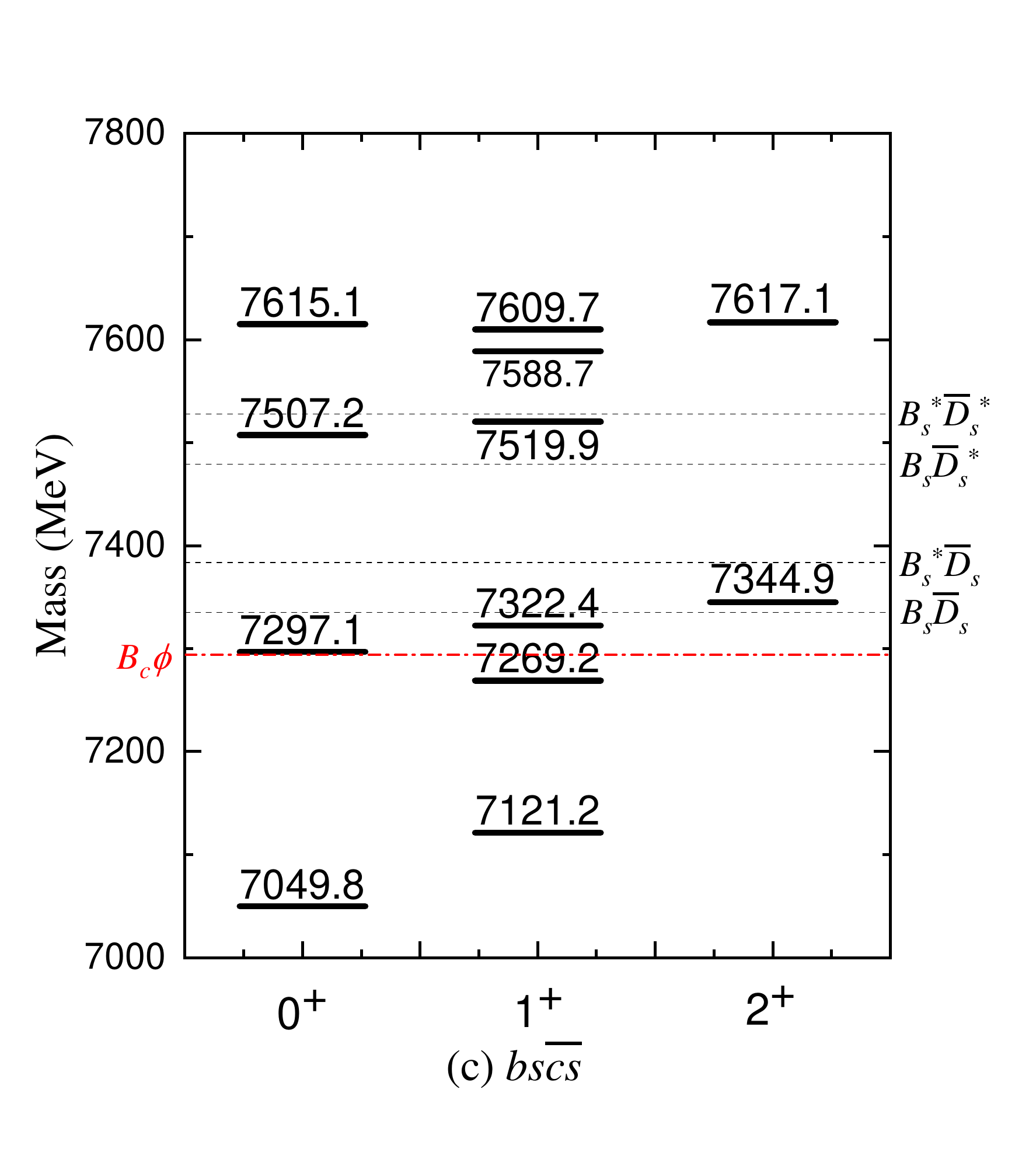}
\end{minipage}

\vspace{-0.5cm}
\caption{\label{fig3}
Mass spectra of $S$-wave tetraquark states (a) $bn\bar{c}\bar{n}$ ($n=u,d$), (b) $bs\bar{c}\bar{n}$, and (c) $bs\bar{c}\bar{s}$ with different quantum numbers.
The black dashed and red dot-dashed lines demonstrate all possible meson-meson thresholds.}
\end{figure*}

The parameters, such as $m_{ij}$ and $v_{ij}$, in the ICMI model can be extracted by fitting the masses, especially the low-lying conventional hadrons, which have been observed in experiments.
In this work, we adopt the parameters obtained in Ref.~\cite{Guo:2021mja}.
Now, we have calculated the mass spectra and wave functions of the tetraquark $Qq\bar{Q}\bar{q}$ ($Q=c,b$, and $q=u,d,s$) systems with various quantum numbers $J^P = 0^{+(+)}$, $1^{+(\pm)}$, and $2^{+(+)}$. Tetraquark state has definite $\mathcal{C}$-parity as long as $Q(q)$ and $\bar Q (\bar q)$ are the same flavor. The $\mathcal{C}$-parity of the tetraquark state is determined by the basis as discussed in the previous section. The tetraquark states can strongly decay into a pair of mesons. The $\mathcal{C}$-parity is conserved in these processes. For the $D\bar D$ ($D^\ast\bar D^\ast$) pair, its $\mathcal{C}$-parity of can be estimated by $(-1)^{L+S}$, where $S$ is the total spin and $L$ is relative angular momentum. While for other meson pairs, such as $D\bar D^\ast$, $D_s\bar D_s^\ast$, the $\mathcal{C}$-parity can be either positive or negative.
The mass spectra of the tetraquark states are shown in Figs.~\ref{fig1}-\ref{fig3}.
Also for comparison, we plot all possible meson-meson thresholds in each figure.
The superposition amplitudes $\{c_i\}$ of the corresponding color-spin wave functions for each tetraquark state are listed in Tables~\ref{atab1}-\ref{atab3}.

%-----------------------------------------------------
%-----------------------------------------------------

%-----------------------------------------------------
%-----------------------------------------------------

%-----------------------------------------------------
%-----------------------------------------------------
%---------------------------------------------------------------------------------------------------------------------
\begin{table*}[!htbp]
	\renewcommand\arraystretch{1.6}
{\tabcolsep0.207in
\begin{tabular}{ccccc}
\toprule[1pt]\toprule[1pt]
System & $J^{PC}$ & Mass & $\{c_i\}$ for $ |(Q_1\bar Q_3)(q_2\bar q_4)\rangle$ basis& $\{c_i\}$ for $ |(Q_1\bar q_4)(q_2\bar Q_3)\rangle$ basis    \\
 \midrule[1pt]
$cn\bar{c}\bar{n}$
&$0^{+(+)}$ & 4146.5& (0.06, -0.16, -0.96, -0.23) &   (-0.21, 0.90, 0.25, 0.29) \\
(a)&     & 3970.9& (0.06, -0.72, -0.05, 0.69) &         (-0.37, -0.19, 0.82, 0.39) \\
&        & 3748.2 & (0.11, 0.68, -0.27, 0.67) &     (-0.86, -0.24, -0.26, -0.38)\\
&        & 3109.3& (-0.99, 0.02, -0.09, 0.10) &     (-0.30, 0.31, -0.44, 0.79)\\
&$1^{+(-)}$ & 4125.2 & (-0.01, -0.06, 0, -0.45, 0.89, 0)&   (0.20, 0.20, -0.88, -0.11, -0.11, 0.35)\\
&        & 4036.0 &(-0.30, -0.08, 0, 0.85, 0.42, 0)&              (0.54, 0.54, 0.24, -0.39, -0.39, -0.25)\\
&        & 3731.7 &(-0.95, 0.04, 0, -0.27, -0.13, 0)&            (-0.34, -0.34, -0.32, -0.36, -0.36, -0.63)\\
&        & 3226.8 &(-0.02, -0.99, 0, -0.06, -0.09, 0)&          (-0.24, -0.24, -0.26, -0.45, -0.45, 0.64)\\
&$1^{+(+)}$& 4002.1 &(0, 0, -0.56, 0, 0, 0.83)&             (0.42, -0.42, 0, -0.57, 0.57, 0) \\
&        & 3813.1 &(0, 0, -0.83, 0, 0, -0.56)&                     (-0.57, 0.57, 0, -0.42, 0.42, 0)\\
&$2^{+(+)}$ &4107.5         & (-0.28, 0.96)  &        (-0.81, 0.58)\\
&        & 3852.2        & (-0.96, -0.28)  &                  (-0.58, -0.81)\\[1em]
\midrule[1pt]
$cs\bar{c}\bar{n}$
&$0^{+}$ & 4220.8& (0.08, -0.19, -0.95, -0.24) &        (0.19, -0.90, -0.29, 0.29)\\
(b)&        & 4078.5& (0.08, -0.73, -0.01, 0.68) &            (0.33, 0.21, -0.83, -0.39) \\
&        & 3857.2& (-0.19, -0.65, 0.29, -0.68) &                (-0.85, -0.25, -0.20, -0.43) \\
&        & 3455.1& (-0.98, 0.05, -0.13, 0.17)  &                   (-0.38, 0.30, -0.43, 0.77) \\
&$1^{+}$ & 4203.2 & (-0.03, 0.07, 0.01, 0.56, -0.83, -0.01)&   (-0.13, -0.12, 0.90, 0.07, 0.06, -0.39)\\
&        & 4133.2 &(0.32, 0.14, -0.01, -0.78, -0.52, 0.01)&          (0.55, 0.53, 0.13, -0.44, -0.43, -0.18)\\
&        & 4105.5 &(-0.001, -0.002, -0.60, 0.02, 0.001, 0.80)&      (0.39, -0.40, 0.01, -0.58, 0.59, -0.003)\\
&        & 3923.8 &(0.02, -0.002, 0.80, -0.001, 0.003, 0.60)&        (0.59, -0.58, 0.01, 0.40, -0.39, 0.02)\\
&        & 3846.6 &(-0.95, 0.09, 0.01, -0.27, -0.15, 0.01)&             (0.33, 0.35,  0.33, 0.33, 0.34, 0.66) \\
&        & 3575.6 &(0.04, 0.99, 0.00, 0.09, 0.14, 0.001)&               (-0.28, -0.28, 0.25, -0.45, -0.45, 0.62)\\
&$2^{+}$ & 4209.2         & (-0.29, 0.96)  &                               (-0.80, -0.60)\\
&        & 3967.2        & (-0.96, -0.29)  &                                   (0.60, -0.80) \\[1em]
\midrule[1pt]
 $cs\bar{c}\bar{s}$ &$0^{++}$ & 4302.7& (0.11, -0.24, -0.94, -0.22) &       (-0.17, 0.88, 0.34, -0.28)\\
(c)&        &4195.1& (0.10, -0.75, 0.05, 0.65) &                  (0.27, 0.25, -0.84, -0.40) \\
&        & 3976.5& (-0.35, -0.60, 0.27, -0.67) &                  (-0.79, -0.29, -0.09, -0.54) \\
&        & 3739.3& (-0.93, 0.11, -0.21, 0.29)  &                      (-0.53, 0.28, -0.41, 0.69) \\
&$1^{+-}$ &4289.2& (0.07, -0.08, 0, -0.65, 0.75, 0) &         (0.05, 0.05, -0.90, -0.02, -0.02, 0.43)\\
&        & 4232.3& (-0.34, -0.22, 0, 0.69, 0.60, 0)&               (0.51, 0.51, 0.03, -0.48, -0.48, -0.10)\\
&        & 3975.2& (0.92, -0.25, 0, 0.27, 0.12, 0)&                 (0.30, 0.30, 0.37, 0.25, 0.25, 0.75)\\
&        & 3864.9& (-0.16, -0.94, 0, -0.18, -0.24, 0)&             (-0.38, -0.38, 0.22, -0.45, -0.45, 0.50)\\
& $1^{++}$ &4215.2& (0, 0, -0.66, 0, 0, 0.75) &                  (0.34, -0.34, 0, -0.62, 0.62, 0) \\
&        &4039.1& (0, 0, -0.75, 0, 0, -0.66) &                       (-0.62, 0.62, 0, -0.34, 0.34, 0)\\
&$2^{++}$ & 4309.7& (-0.33, 0.94)  &                               (-0.78, -0.63)\\
&        & 4092.6& (-0.94, -0.33)   &                                    (0.63, -0.78)  \\
\bottomrule[1pt]\bottomrule[1pt]
\end{tabular}}
\caption{\label{atab1} Masses (in MeV) and  color-spin wave functions (represented by the superposition amplitudes $\{c_i\}$) of the $S$-wave hidden-charm tetraquark states with quantum numbers
$J^{PC} = 0^{+(+)}$, $1^{+(\pm)}$, and $2^{+(+)}$. The label (a), (b), and (c) are corresponding to Fig.~\ref{fig1}.}
\end{table*}
%---------------------------------------------------------------------------------------------------------------------

%---------------------------------------------------------------------------------------------------------------------
\begin{table*}[!htbp]
	\renewcommand\arraystretch{1.6}
{\tabcolsep0.207in
\begin{tabular}{ccccc}
\toprule[1pt]\toprule[1pt]
System & $J^{PC}$ & Mass & $\{c_i\}$ for $ |(Q_1\bar Q_3)(q_2\bar q_4)\rangle$ basis& $\{c_i\}$ for $ |(Q_1\bar q_4)(q_2\bar Q_3)\rangle$ basis    \\
\midrule[1pt]
 $bn\bar{b}\bar{n}$ &$0^{+(+)}$ & 10821.9& (-0.05, 0.01, 0.98, 0.18) & (0.30, -0.88, -0.15, 0.35)\\
 (a) &                   &  10688.5& (0.03, -0.17, -0.18, 0.97) & (-0.82, -0.29, 0.47, 0.17)\\
     &                     &  10221.4& (0.01, 0.98, -0.04, 0.17) & (-0.44, -0.21, -0.74, -0.46)\\
     &                     &  9533.9& (0.99, -0.01, 0.06, -0.02) & (0.21, -0.32, 0.46, -0.80) \\
     &$1^{+(-)}$& 10812.2 & (0.002, -0.06, 0, -0.20, 0.98, 0)& (0.36, 0.36, -0.77, -0.16, -0.16, 0.32)\\
     &        & 10739.2 &(-0.13, -0.02, 0, 0.97, 0.20, 0)&(0.53, 0.53, 0.49, -0.27, -0.27, -0.25)\\
     &        & 10164.9 &(0.99, -0.01, 0, 0.12, 0.02, 0)&(0.23, 0.23, 0.30, 0.44, 0.44, 0.64)\\
     &          & 9595.6 &(-0.003, -0.998, 0, -0.01, -0.06, 0)&(-0.20, -0.20, 0.27, -0.46, -0.46, 0.65)\\
     &       $1^{+(+)}$ & 10713.7 &(0, 0, -0.15, 0, 0, 0.99)&(0.62, -0.62, 0, -0.33, 0.33, 0) \\
     &               & 10224.2 &(0, 0, -0.99, 0, 0, -0.15)&(-0.33, 0.33, 0, -0.62, 0.62, 0)\\
     &$2^{+(+)}$& 10757.2  & (-0.14, 0.99)& (-0.89, -0.45)\\
     &       & 10226.9        & (-0.99, -0.14)& (0.45, -0.89)\\[1em]
\midrule[1pt]
 $bs\bar{b}\bar{n}$ &$0^{+}$ & 10883.6& (0.07, -0.01, -0.97, -0.23)& (-0.26, 0.88, 0.13, -0.37)\\
 (b) &             & 10776.5& (0.04, -0.19, -0.22, 0.96)& (-0.82, -0.25, 0.49, 0.15) \\
&                    & 10335.7& (0.02, 0.98, -0.05, 0.19)& (-0.46, -0.22, -0.73, -0.46) \\
&                    & 9886.2& (0.99, -0.01, 0.08, -0.03) & (0.23, -0.33, 0.46, -0.79)\\
&$1^{+}$& 10874.7& (0.02, -0.07, -0.001, -0.31, 0.95, 0.01)& (-0.30, -0.28, 0.82, 0.13, 0.13, -0.36)\\
&       & 10827.2& (0.14, 0.04, 0.01, -0.94, -0.31, -0.04)& (0.59, 0.54, 0.40, -0.30, -0.28, -0.21)\\
&       & 10803.1& (0.01, 0.002, -0.17, -0.03, -0.02, 0.99)& (0.59, -0.64, -0.01, -0.33, 0.36, 0.01)\\
&       & 10338.9& (0.004, 0, 0.99, 0, -0.001, 0.16)& (0.34, -0.34, 0.001, 0.62, -0.62, 0.003)\\
&       & 10280.1& (0.99, -0.01, -0.003, 0.13, 0.03, -0.002)& (-0.24, -0.24, -0.31, -0.43, -0.44, -0.64) \\
&       & 9947.9& (-0.01, -0.99, 0, -0.02, -0.08, 0)& (-0.21, -0.21, 0.28, -0.46, -0.46, 0.65)\\
&$2^{+}$& 10847.8& (-0.14, 0.99)& (-0.89, -0.46)\\
&       & 10342.2& (-0.99, -0.14) & (0.46, -0.89)\\[1em]
\midrule[1pt]
$bs\bar{b}\bar{s}$ &$0^{++}$ & 10952.6& (0.09, 0.001, -0.96, -0.27)& (-0.22, 0.89, 0.12, -0.39)\\
(c) &               & 10864.3& (0.06, -0.22, -0.26, 0.94) & (0.82, 0.21, -0.52, -0.13)\\
&                    & 10460.5& (0.03, 0.98, -0.06, 0.21) & (-0.47, -0.23, -0.71, -0.47) \\
&                     & 10185.2& (0.99, -0.02, 0.10, -0.04) & (-0.25, 0.35, -0.46, 0.78) \\
&$1^{+-}$& 10945.9 & (0.05, -0.09, 0, -0.48, 0.87, 0)& (0.18, 0.18, -0.88, -0.08, -0.08, 0.41)\\
&           & 10913.0& (-0.15, -0.07, 0, 0.86, 0.48, 0)& (0.60, 0.60, 0.24, -0.33, -0.33, -0.14)\\
&           & 10406.3& (0.99, -0.03, 0, 0.15, 0.03, 0)& (0.24, 0.24, 0.32, 0.43, 0.43, 0.65)\\
&           & 10246.7& (-0.02, -0.99, 0, -0.03, -0.11, 0)& (-0.23, -0.23, 0.29, -0.46, -0.46, 0.63)\\
&$1^{++}$       & 10891.9& (0, 0, -0.19, 0, 0, 0.98) & (0.61, -0.61, 0, -0.36, 0.36, 0) \\
&                      & 10464.4& (0, 0, -0.98, 0, 0, -0.19)& (-0.36, 0.36, 0, -0.61, 0.61, 0)\\
&$2^{++}$& 10937.3& (-0.16, 0.98)& (-0.88, 0.48)\\
&                & 10468.3& (-0.98, -0.16) & (-0.48, -0.88)\\
\bottomrule[1pt]\bottomrule[1pt]
\end{tabular}}
\caption{\label{atab2} Masses (in MeV) and  color-spin wave functions (represented by the superposition amplitudes $\{c_i\}$) of the $S$-wave hidden-bottom tetraquark states with quantum numbers
$J^{PC} = 0^{+(+)}$, $1^{+(\pm)}$, and $2^{+(+)}$. The label (a), (b), and (c) are corresponding to Fig.~\ref{fig2}.}
\end{table*}
%---------------------------------------------------------------------------------------------------------------------

%---------------------------------------------------------------------------------------------------------------------
\begin{table*}[!htbp]
	\renewcommand\arraystretch{1.6}
{\tabcolsep0.24in
\begin{tabular}{ccccc}
\toprule[1pt]\toprule[1pt]
System & $J^{P}$ & Mass & $\{c_i\}$ for $ |(Q_1\bar Q_3)(q_2\bar q_4)\rangle$ basis & $\{c_i\}$ for $ |(Q_1\bar q_4)(q_2\bar Q_3)\rangle$   basis  \\
\midrule[1.2pt]
$bn\bar{c}\bar{n}$
&$0^{+}$ & 7471.9& (-0.06, 0.07, 0.97, 0.23) & (0.24, -0.90, -0.18, 0.33) \\
(a)&     & 7303.8& (0.06, -0.46, -0.17, 0.87)  & (-0.65, -0.21, 0.68, 0.26)  \\
&        & 7069.5 & (0.05, 0.89, -0.16, 0.43) & (-0.68, -0.23, -0.55, -0.43) \\
&        & 6406.2& (0.99, -0.02, 0.08, -0.06)& (0.25, -0.32, 0.45, -0.80)\\
&$1^{+}$ & 7461.8 & (-0.01, -0.06, 0.02, -0.27, 0.95, 0.14)& (0.41, 0.21, -0.80, -0.17, -0.13, 0.32)\\
&        & 7401.3 &(-0.18, -0.03, -0.12, 0.83, 0.16, 0.49)&(0.73, 0.13, 0.41, -0.46, -0.07, -0.26)\\
&        & 7322.6 &(0.15, 0.05, -0.35, -0.43, -0.23, 0.79)&(0.17, -0.72, -0.11, -0.21, 0.62, 0.12)\\
&        & 7088.7 &(0.14, 0.02, 0.93, -0.03, -0.07, 0.34) &(0.43, -0.47, 0.06, 0.63, -0.44, 0.07)\\
&        & 7030.0 &(0.96, -0.03, -0.10, 0.23, 0.09, -0.08) &(0.23, 0.38, 0.33, 0.34, 0.44, 0.63) \\
&        & 6477.1 &(0.01, 0.99, -0.01, 0.04, 0.08, -0.03) &(0.20, 0.24, -0.26, 0.46, 0.45, -0.65)\\
&$2^{+}$ & 7424.9         & (-0.22, 0.98)    & (-0.85, -0.53)  \\
&        & 7105.1        & (-0.98, -0.22)   & (0.53, -0.85)  \\[1em]
\midrule[1pt]
$bs\bar{c}\bar{n}$
&$0^{+}$ & 7544.2& (-0.08, 0.07, 0.96, 0.26)& (0.21, -0.90, -0.18, 0.35)\\
(b)&     & 7404.5& (0.08, -0.49, -0.19, 0.85) & (-0.63, -0.19, 0.72, 0.25) \\
&        & 7182.0& (0.09, 0.87, -0.18, 0.45) & (-0.69, -0.23, -0.51, -0.46) \\
&        & 6756.7& (0.99, -0.03, 0.10, -0.09)  & (0.29, -0.32, 0.45, -0.78)  \\
&$1^{+}$ & 7535.8 &(-0.01, 0.08, -0.02, 0.38, -0.91, -0.13)&(-0.33, -0.15, 0.84, 0.14, 0.10, -0.36)\\
&        & 7499.5 &(-0.19, -0.05, -0.12, 0.79, 0.25, 0.52)&(0.77, 0.14, 0.32, -0.49, -0.08, -0.22)\\
&        & 7421.1 &(0.17, 0.08, -0.37, -0.42, -0.27, 0.76)&(0.13, -0.71, -0.08, -0.19, 0.66, 0.09)\\
&        & 7203.3 &(0.16, 0.04, 0.91, -0.03, -0.08, 0.36)&(0.44, -0.47, 0.07, 0.64, -0.41, 0.07)\\
&        & 7145.4 &(0.96, -0.05, -0.12, 0.23, 0.10, -0.09) &(0.22, 0.40, 0.33, 0.32, 0.43, 0.64) \\
&        & 6827.9 &(0.02, 0.99, -0.02, 0.05, 0.11, -0.05)&(0.21, 0.28, -0.27, 0.45, 0.45, -0.63)\\
&$2^{+}$ & 7524.4         & (-0.23, 0.97)    & (-0.84, -0.54)  \\
&        & 7221.2        & (-0.97, -0.23)   & (0.54, -0.84) \\[1em]
\midrule[1pt]
$bs\bar{c}\bar{s}$
&$0^{+}$ & 7615.1& (-0.11, 0.08, 0.95, 0.27) & (0.19, -0.89, -0.20, 0.37) \\
(c)&     & 7507.2& (0.12, -0.56, -0.17, 0.80) & (-0.55, -0.18, 0.77, 0.26) \\
&        & 7297.1& (0.17, 0.82, -0.20, 0.51)  & (-0.72, -0.27, -0.41, -0.50)  \\
&        & 7049.8& (0.97, -0.07, 0.16, -0.16)   & (0.39, -0.33, 0.44, -0.74)   \\
&$1^{+}$ & 7609.7 & (-0.07, 0.10, -0.05, 0.56, -0.82, -0.03) & (-0.15, -0.08, 0.88, 0.04, 0.08, -0.44)\\
&        & 7588.7 &(-0.21, -0.09, -0.15, 0.66, 0.44, 0.54)&(0.80, 0.15, 0.12, -0.55, -0.10, -0.13)\\
&        & 7519.9 &(0.19, 0.12, -0.43, -0.41, -0.29, 0.72)&(0.10, -0.65, -0.07, -0.19, 0.72, 0.07)\\
&        & 7322.4 &(0.19, 0.08, 0.88, -0.02, -0.08, 0.42) &(0.49, -0.49, 0.07, 0.63, -0.34, 0.06)\\
&        & 7269.2 &(0.94, -0.13, -0.13, 0.26, 0.10, -0.11) &(0.20, 0.41, 0.36, 0.26, 0.38, 0.67) \\
&        & 7121.2 &(0.07, 0.97, -0.05, 0.10, 0.18, -0.09)&(0.24, 0.38, -0.27, 0.44, 0.46, -0.58)\\
&$2^{+}$ & 7617.1         & (-0.27, 0.96)  & (-0.82, -0.57)  \\
&        & 7344.9        & (-0.96, -0.27) & (0.57, -0.82)  \\
\bottomrule[1pt]\bottomrule[1pt]
\end{tabular}}
\caption{\label{atab3} Masses (in MeV) and  color-spin wave functions (represented by the superposition amplitudes $\{c_i\}$) of the $S$-wave mixed-charm-bottom tetraquark states with quantum numbers
$J^{P} = 0^{+}$, $1^{+}$, and $2^{+}$. The label (a), (b), and (c) are corresponding to Fig.~\ref{fig3}.}
\end{table*}
%---------------------------------------------------------------------------------------------------------------------

\subsection{Hidden-charm tetraquark states with $cn\bar{c}\bar{n}$, $cs\bar{c}\bar{n}$, and $cs\bar{c}\bar{s}$ configurations }

\subsubsection{The $cn\bar{c}\bar{n}$ configuration }

The mass spectra of the tetraquark states with $cn\bar{c}\bar{n}$ configuration are shown in Fig.~\ref{fig1}(a). The superposition amplitudes of the corresponding wave functions are listed in Table~\ref{atab1}(a).
It is worth noting that the notation $n$ represents a $u$ or $d$ quark from now on.
And in our ICMI model, $u$ and $d$ quarks have the same mass.

(1). There are four $S$-wave tetraquark states with $J^{P}=0^{+}$ ($J^{PC}=0^{++}$ for charge-neutral states, such as $[cu\bar{c}\bar{u}]$ and $[cd\bar{c}\bar{d}]$).
Their masses are 3109.3 MeV, 3748.2 MeV, 3970.9 MeV, and 4146.5 MeV, respectively.
The lowest state, 3109.3 MeV, lies slightly below the $\eta_c \pi$ threshold and its wave function has a large fraction on the $\eta_c \pi$ basis ($c_1=-0.99$). The 3748.2 MeV is above the $\eta_c \pi$ and $D\bar D$ thresholds. Considering the angular momentum and $\mathcal{C}$-parity conservation (for charge-neutral states), this state is allowed to decay strongly into the mesons $\eta_c \pi$ or $D\bar D$. In this paper, we only discuss the $S$-wave decay and neglect the high-order decays such as $P$ and $D$-wave, which usually give a small contribution~\cite{Weng:2019ynv}. The 3970.9 MeV is above the $\eta_c \pi$, $D\bar D$, and $J/\psi \rho$ thresholds.
Therefore, compared with 3748.2 MeV, this state has one more possible decay mode, $J/\psi \rho$.
The remaining state with a large mass, 4146.5 MeV, can decay into mesons $\eta_c \pi$, $D\bar D$, $J/\psi \rho$, and $D^\ast \bar {D}^\ast$.

(2). There are six $S$-wave tetraquark states with $J^{P}=1^{+}$. For charge-neutral tetraquark states, they can be separated into two classes based on the $\mathcal{C}$-parity. There are four $S$-wave tetraquark states with $J^{PC}=1^{+-}$.
The lowest state with a mass of 3226.8 MeV, is slightly lower than the $J/\psi \pi$ threshold. Its wave function has a large fraction on the $J/\psi \pi$ basis ($c_2=-0.99$). It can decay into $J/\psi$ and $\pi$ easily due to the large fraction and small mass difference.
Similarly, the state with a mass of 3731.7 MeV is close to the $\eta_c \rho$ threshold. The large fraction $c_i$ indicates it is can decay into $\eta_c$ and $\rho$ easily. In the meantime, it can also decay into the mesons $J/\psi \pi$ (decay into $D\bar D$ via $S$-wave breaks the $\mathcal{C}$-parity). The other two states, 4036.0 and 4125.2 MeV lie above all meson-meson thresholds. Considering the conservation law in the decay process, the possible decay channels are $J/\psi \pi$, $\eta_c \rho$, and $D\bar {D}^{\ast}$.
Another two $S$-wave tetraquark states with positive $\mathcal{C}$-parity, $J^{PC}=1^{++}$.
Their masses are 3813.1 MeV and 4002.1 MeV.
The state with 3813.1 MeV is near the $J/\psi \rho$ and $D \bar {D}^\ast$ threshold.
Thus the state can decay into mesons $\pi^+ \pi^- J/\psi$ via the quantum off-shell decay process $\rho \rightarrow \pi^+\pi^-$. The tetraquark states $Qq\bar Q\bar q$ have no isospin symmetry. However, if considering the mixing between different tetraquark states, such as $cu\bar c\bar u$ and $cu\bar d\bar d$ can form a state with definite isospin. Especially, the state $|X\rangle=(|cu\bar c\bar u\rangle - |cu\bar d\bar d\rangle)/\sqrt{2}$ with $I=1$ is in good agreement with the experimental results of $X(3872)$ with $I(J^{PC})=0(1^{++})$~\cite{Belle:2003nnu}. However, our predicted mass of 3813.1 MeV is almost 60 MeV lower than 3872 MeV. If that is true, $X(3872)$ probably is a mixing state of excited charmonium ($\chi_{c1}(2P)$) and tetraquark $|X\rangle$ states, which is similar to a mixed molecule-charmonium state~\cite{Badalian:2012jz,Matheus:2009vq}.
Our results are also consistent with the conclusions of some previous theoretical studies~\cite{Maiani:2004vq,Hogaasen:2005jv}.
The 4002.1 MeV is slightly below the $D^{\ast}{\bar D}^{\ast}$ threshold. It can decay into $D^{\ast}\bar {D}^{\ast}$ via the off-shell process. Therefore, all possible decay channels are $D^{\ast}{\bar D}^{\ast}$, $D\bar D$, $J/\psi \rho$, and $\eta_c \pi$.

(3). There are two $S$-wave tetraquark states with $J^{P}=2^{+}$ ($J^{PC}=2^{++}$ for charge-neutral states), which masses are 3852.2 MeV and 4107.5 MeV, respectively.
The 3852.2 MeV is very close to the $J/\psi \rho$ threshold and its wave function has a large component of $J/\psi \rho$ basis ($c_1=0.96$).  The 4107.5 MeV is above $D^{\ast} \bar {D}^{\ast}$ and $J/\psi \rho$ thresholds and may be allowed decay into $D^{\ast} \bar {D}^{\ast}$ and $J/\psi \rho$.

\subsubsection{The $cs\bar{c}\bar{n}$ configuration }

The mass spectra of the tetraquark states with $cs\bar{c}\bar{n}$ configuration are shown in Fig.~\ref{fig1}(b).
For comparison, we also show the corresponding meson-meson thresholds in this figure.
The masses and wave functions amplitudes $\{c_i\}$ are listed in Table~\ref{atab1}(b).
For the $cs\bar{c}\bar{n}$ state, the $\mathcal{C}$-parity is not a good quantum number. So, the quantum numbers of the $S$-wave tetraquark state $cs\bar{c}\bar{n}$ are $J^{P}=0^{+}$, $1^{+}$, and $2^{+}$.

(1). There are four $J^{P}=0^{+}$ states, which masses are 3455.1 MeV, 3857.2 MeV, 4078.5 MeV, and 4220.8 MeV.
The state with the smallest mass, 3455.1 MeV, is close to the $\eta_c K$ threshold and its wave function has a large component of $\eta_c K$ basis ($c_1=0.98$). So, it can decay into $\eta_cK$ easily.
The 3857.2 MeV is above the $\eta_c K$ and $D\bar D_s$ thresholds and can naturally decay into mesons $\eta_c  K$ and $D\bar D_s$.
The 4078.5 MeV is below the $D^{\ast}\bar {D}_s^{\ast}$ threshold but above other possible meson-meson thresholds.
The possible decay channels are $\eta_c K$, $D\bar D_s$, and $J/\psi K^\ast$.
The state with the largest mass 4220.8 MeV lies above all possible meson-meson thresholds, which can decay to $\eta_c K$, $D\bar {D}_s$, $J/\psi K^\ast$, and $D^{\ast} \bar {D}_s^{\ast}$.

(2). There are six $S$-wave tetraquark states with $J^{P}=1^{+}$.
Their masses are 3575.6 MeV, 3846.6 MeV, 3923.8 MeV, 4105.5 MeV, 4133.2 MeV, and 4203.2 MeV, respectively.
The state with the smallest mass, 3575.6 MeV, which wave function has a large component of $J/\psi K$ basis ($c_2=0.99$). Because its mass is very close to the $J/\psi K$ threshold, it can decay into $J/\psi K$ with a large probability.
Similarly, the 3846.6 MeV can decay into $\eta_cK$ pair due to the large fraction $c_1=0.95$.
The 3923.8 MeV is above the $J/\psi K$ and $\eta_c K^\ast$ thresholds and close to the $D^{\ast}\bar {D}_s$ and $D\bar {D}_s^{\ast}$ thresholds.
We can see its wave function has a large $|\beta_3\rangle$ component ($c_3 = 0.80$), indicating that the mesons $\eta_c K^\ast$ component occupies a large proportion.
To study the weights of this state to $D^{\ast} \bar {D}_s$  and $D\bar {D}_s^{\ast}$ bases, we convert the tetraquark configuration $|(Q_1\bar{Q}_3)(q_2\bar{q}_4)\rangle$ to $|(Q_1\bar{q}_4)(q_2\bar{Q}_3)\rangle$.
As shown in Table~\ref{atab1}, we can see the amplitudes of the state (3923.8 MeV) under this set of $|(Q_1\bar{q}_4)(q_2\bar{Q}_3)\rangle$ basis is $(0.59, -0.58, 0.01, 0.4, -0.39, 0.02)$.
It indicates that the first two components are approximately the same and relatively large, indicating that the tetraquark state can be allowed to decay into the mesons $D^{\ast} \bar {D}_s$ and $D\bar {D}_s^{\ast}$.
The results nicely explain the nature of exotic resonant structure $Z_{cs}(3985)^-$~\cite{BESIII:2020qkh}. However, another exotic state $Z_{cs}(4000)^+$, whose mass is very close to $Z_{cs}(3985)^-$ but with a larger decay width can not be classified in our calculations.
The 4105.5 MeV is only below the $D^{\ast} \bar {D}_s^{\ast}$ threshold and thus can naturally decay into the mesons $J/\psi  K$, $\eta_c  K^\ast$, $J/\psi  K^\ast$, $D^{\ast} \bar {D}_s$, and $D\bar {D}_s^{\ast}$.
The remaining two states 4133.2 MeV and 4203.2 MeV lie above all possible meson-meson thresholds so that all decay modes are possible.
In addition, it is worth noting that the mass and decay channel of the tetraquark state 4203.2 MeV with quark content $cu\bar c\bar s$ is probably the experimentally observed $Z_{cs}(4220)^+$~\cite{LHCb:2021uow}.

(3). For $J^{P}=2^{+}$, masses of two $S$-wave tetraquark states are 3967.2 MeV and 4209.2 MeV.
The lower state lies slightly below the $J/\psi K^\ast$ threshold and its wave function has a large component of $J/\psi K^\ast$ basis ($c_1=-0.96$). So, it can be searched in the $J/\psi K^\ast$ channel.
The higher state lies above the $J/\psi K^\ast$ and $D^{\ast}\bar {D}_s^{\ast}$ thresholds and its dominant decay modes are mesons $J/\psi  K^\ast$ and $D^{\ast} \bar {D}_s^{\ast}$.

\subsubsection{The $cs\bar{c}\bar{s}$ configuration }

For the tetraquark $cs\bar{c}\bar{s}$ configuration, it has definite $\mathcal{C}$-parity.
All possible quantum numbers are $J^{PC}=0^{++}$, $1^{+-}$, $1^{++}$, and $2^{++}$.
The mass spectra of the tetraquark state with $cs\bar{c}\bar{s}$ configuration are shown in Fig.~\ref{fig1}(c) with corresponding meson-meson thresholds. For the $s\bar s$ system, there is no pure spin singlet state due to the mixing between $u\bar u$ and $d\bar d$. For the spin-triplet state, the mixing angle is opportune to form a very nearly pure $s\bar s$ state, which is named $\phi$ meson. So, we only plot the $\eta_c\phi$ and $J/\psi\phi$ thresholds.
The superposition amplitudes $\{ c_i \}$ of the corresponding tetraquark wave functions are listed in Table~\ref{atab1}(c).

(1). There are four $S$-wave tetraquark states with $J^{PC}=0^{++}$, whose masses are 3739.3 MeV, 3976.5 MeV, 4195.1 MeV, and 4302.7 MeV, respectively.
From the wave function of the lowest state, 3739.3 MeV, we can see it has a large component based on $\eta_c$ and the spin singlet state $s\bar s$. Because $s\bar s$ content mixed with $u\bar u$ and $d\bar d$. So, this state may be a superposition state of $\eta_c \eta$ and $\eta_c\eta'$.
We will not discuss the decay channels of these mixed states in this paper.
Considering its mass is largely below the meson-meson thresholds, it would be a stable tetraquark state.
The 3976.5 MeV is close to and above the $D_s\bar {D}_s$ threshold and thus can naturally decay into mesons $D_s\bar {D}_s$.
Now, we convert the $|(Q_1\bar{Q}_3)(q_2\bar{q}_4)\rangle$ configuration to the $|(Q_1\bar{q}_4)(q_2\bar{Q}_3)\rangle$ configuration. Under this set of color-spin basis, the superposition amplitude $\{ c_i \}$ of this tetraquark state is $(-0.79, -0.29, -0.09, -0.54)$.
It can be found that the amplitude $c_1$ of the $D_s\bar {D}_s$ component is -0.79. So, the main decay channel of this state is mesons $D_s\bar {D}_s$.
These analyses indicate this state probably is the newly observed exotic hadron state $X(3960)$ with $J^{PC}=0^{++}$.
The 4195.1 MeV is above the $D_s\bar {D}_s$ and $J/\psi \phi$ thresholds but below the $D_s^{\ast}\bar {D}_s^{\ast}$ threshold. So, the possible decay channels are $D_s\bar {D}_s$ and $J/\psi  \phi$. And this state might be the observed $X_0(4140)$ at LHCb~\cite{LHCb:2022vsv}.
The last state with a mass of 4302.7 MeV is above all possible meson-meson thresholds and can be allowed to decay into $D_s  \bar{D}_s$, $D_s^{\ast}\bar{D}_s^{\ast}$, and $J/\psi  \phi$.

(2). There are four $S$-wave tetraquark states with $J^{PC}=1^{+-}$ and masses 3864.9 MeV, 3975.2 MeV, 4232.3 MeV, and 4289.2 MeV, respectively.
Without the spin singlet $s\bar s$ meson, the lowest state with a mass of 3864.9 MeV is probably a stable tetraquark state.
While the state, 3975.2 MeV, with a large $\eta_c\phi$ component, could decay into $\eta_c\phi$ via the off-shell process.
The two highest states, 4232.3 MeV and 4289.2 MeV are located above all possible meson-meson thresholds. Considering the conservation of angular momentum and $\mathcal{C}$-parity in the decay process, they can decay into the $D_s^{\ast}  \bar{D}_s^{\ast}$ and $D_s\bar{D}_s^{\ast}$ via the $S$ wave. Because $D_s\bar{D}_s^{\ast}$ pair can form a negative $\mathcal{C}$-parity state via, $(D_s\bar{D}_s^{\ast}-{D}_s^{\ast}\bar D_s)/\sqrt{2}$.
Besides, there are two $S$-wave tetraquark states with $J^{PC}=1^{++}$ and masses 4039.1 MeV and 4215.2 MeV, respectively.
The state with a mass of 4039.1 MeV has large fractions on both $J/\psi \phi$ and $D_s\bar{D}_s^{\ast}$ bases. Satisfying the $\mathcal{C}$-parity conversion, it can decay into $J/\psi \phi$ and a mixed state of $D_s\bar{D}_s^{\ast}$ and its antiparticle, $(D_s\bar{D}_s^{\ast}+D_s^{\ast}\bar D_s)/\sqrt{2}$. The state, 4215.2 MeV, is very close to the $D_s^{\ast} \bar{D}_s^{\ast}$ threshold.
It can decay into $J/\psi \phi$, while the $P$ wave decay into the final $D_s^{\ast} \bar{D}_s^{\ast}$ pair is allowed.

(3). For $J^{PC}=2^{++}$, we find two $S$-wave tetraquark states with masses of 4092.6 MeV and 4309.7 MeV, respectively.
The wave function of 4092.6 MeV has a large $J/\psi \phi$ component, it can decay into $J/\psi \phi$.
The 4309.7 MeV is above all possible meson-meson thresholds and can be allowed the decay into mesons $J/\psi  \phi$ and $D_s^{\ast}  \bar{D}_s^{\ast}$.

\subsection{Hidden-bottom tetraquark states with $bn\bar{b}\bar{n}$, $bs\bar{b}\bar{n}$, and $bs\bar{b}\bar{s}$ configurations }

The hidden-bottom tetraquark states can be realized by replacing the charm quark in the hidden-charm tetraquark states with the bottom quark.
Substituting the corresponding model parameters into Eq.~(\ref{Ha1}), we obtain the mass spectra and wave functions of the $S$-wave hidden-bottom tetraquark states.
These hidden-bottom tetraquark states can be divided into three configurations, namely $bn\bar{b}\bar{n}$, $bs\bar{b}\bar{n}$, and $bs\bar{b}\bar{s}$.
Their mass spectra and corresponding meson-meson thresholds are presented in Fig.~\ref{fig2}.
The superposition amplitudes $\{c_i\}$ of the wave functions of these tetraquark states are listed in Table~\ref{atab2}.
The analysis method is similar to the previous hidden-charm tetraquark states. So, we will not discuss them one by one and only focus on two experimental observed states, $Z_b(10610)^+$ and $Z_b(10652)^+$~\cite{Belle:2011aa}. They are observed in the decay channels of $\Upsilon(ns) \pi$ and $h_b(mp)\pi$ with $n=1,2,3$ and $m=1,2$. Their quantum numbers are probably $J^{P}=1^+$ but without the information of the $\mathcal{C}$-parity. In our model, the state with a mass of 10713.7 MeV and 10739.2 MeV with $J^{P}=1^+$ are probably the observed $Z_b(10610)^+$ and $Z_b(10652)^+$ respectively, as shown in Fig.~\ref{fig2}(a). If so, these two states should have different $\mathcal{C}$-parities. Their wave function has a very large fraction on the $\beta_4$ and $\beta_6$ basis, which indicates these two states might be the diquark-antiquark bound states.

%---------------------------------------------------------------------------------------------------------------------
\begin{table}[!htbp]
	\renewcommand\arraystretch{1.6}
{\tabcolsep0.217in
\begin{tabular}{cccc}
\toprule[1pt]\toprule[1pt]
System & $J^{P}$ & Mass & Meson-Meson   \\
\midrule[1.2pt]
$cn\bar{c}\bar{n}$ & $0^{+}$   & 3109.3& $\eta_c\pi$ \\
$cn\bar{c}\bar{n}$&  $1^{+}$   & 3226.8 & $J/\psi \pi$  \\
$cn\bar{c}\bar{n}$&  $1^{+}$     & 3731.7 & $\eta_c \rho$ \\
$cn\bar{c}\bar{n}$&   $2^{+}$     & 3852.2 & $J/\psi \rho$\\
$cs\bar{c}\bar{n}$& $0^{+}$        & 3455.1 & $\eta_c K$\\
$cs\bar{c}\bar{n}$& $1^{+}$       & 3575.6 &$J/\psi K$\\
$cs\bar{c}\bar{n}$&  $1^{+}$      & 3846.6 &$\eta_c K^{\ast}$\\
$cs\bar{c}\bar{n}$&  $2^{+}$      & 3967.2 &$J/\psi K^{\ast}$\\[1em]
%\midrule[1pt]
$bn\bar{b}\bar{n}$ & $0^{+}$   & 9533.9& $\eta_b\pi$ \\
$bn\bar{b}\bar{n}$&  $1^{+}$   & 9595.6 & $\Upsilon \pi$  \\
$bn\bar{b}\bar{n}$&  $1^{+}$     & 10164.9 & $\eta_b \rho$ \\
$bn\bar{b}\bar{n}$&   $1^{+}$     & 10224.2 & $\Upsilon \rho$\\
$bn\bar{b}\bar{n}$&   $2^{+}$     & 10226.9 & $\Upsilon \rho$\\
$bs\bar{b}\bar{n}$& $0^{+}$        & 9886.2 & $\eta_b K$\\
$bs\bar{b}\bar{n}$& $1^{+}$       & 9947.9 &$\Upsilon K$\\
$bs\bar{b}\bar{n}$&  $1^{+}$      & 10280.1 &$\eta_b K^{\ast}$\\
$bs\bar{b}\bar{n}$&  $1^{+}$      & 10338.9 &$\Upsilon K^{\ast}$\\
$bs\bar{b}\bar{n}$&  $2^{+}$      & 10342.2 &$\Upsilon K^{\ast}$\\
$bs\bar{b}\bar{s}$&   $1^{+}$     & 10406.3 &$\eta_b \phi$ \\
$bs\bar{b}\bar{s}$&   $1^{+}$     & 10464.4 &$\Upsilon \phi$\\
$bs\bar{b}\bar{s}$&   $2^{+}$     & 10468.3 &$\Upsilon \phi$\\[1em]
%\midrule[1pt]
$bn\bar{c}\bar{n}$ & $0^{+}$   & 6406.2 & $B_c\pi$ \\
$bn\bar{c}\bar{n}$&  $1^{+}$   & 6477.1 & $B_c^{\ast} \pi$  \\
$bn\bar{c}\bar{n}$&  $1^{+}$     & 7030.0 & $B_c \rho$ \\
$bn\bar{c}\bar{n}$&   $2^{+}$     & 7105.1 & $B_c^{\ast} \rho$\\
$bs\bar{c}\bar{n}$& $0^{+}$        & 6756.7 & $B_c K$\\
$bs\bar{c}\bar{n}$& $1^{+}$       & 6827.9 &$B_c^{\ast} K$\\
$bs\bar{c}\bar{n}$&  $1^{+}$      & 7145.4 &$B_c K^{\ast}$\\
$bs\bar{c}\bar{n}$&  $2^{+}$      & 7221.2 &$B_c^{\ast} K^{\ast}$\\
$bs\bar{c}\bar{s}$&   $2^{+}$     & 7344.9 &$B_c^{\ast} \phi$\\
\bottomrule[1pt]\bottomrule[1pt]
\end{tabular}}
\caption{\label{atab4} Tetraquark states, which have a large overlap ($|c_i|^2>90\%$) to the $Q\bar Q$ and $q\bar q$ mesons. Their masses (in MeV), quantum numbers, and meson-meson contents are listed. $B_c^{\ast}$ with $J^P=1^-$ has not been found in experiments.}
\end{table}
%---------------------------------------------------------------------------------------------------------------------

\subsection{Tetraquark states with $bn\bar{c}\bar{n}$, $bs\bar{c}\bar{n}$, and $bs\bar{c}\bar{s}$ configurations }

The mixed charm-bottom tetraquark state can be obtained by substituting a charm quark in the hidden-charm tetraquark state with a bottom quark, or by substituting a bottom quark in the hidden-bottom tetraquark state with a charm quark.
The difference, however, is that these configurations break the charge conjugation symmetry. Hence, the quantum numbers for the charm-bottom tetraquark systems, $bn\bar{c}\bar{n}$, $bs\bar{c}\bar{n}$, and $bs\bar{c}\bar{s}$, are $J^P=0^+$, $1^+$, and $2^+$.
Similarly, we can obtain their masses and wave functions by using the ICMI model.
Our theoretical results are shown in Fig.~\ref{fig3} and Table~\ref{atab3}. The analysis method is similar to the previous hidden-charm tetraquark states.

\section{summary}
\label{sec.summary}

In this work, we complete a systematic study of the $S$-wave tetraquark states $Qq\bar{Q}\bar{q}$ ($Q=c,b$ and $q=u,d,s$) via the ICMI model, which includes both the chromomagnetic and chromoelectric interaction. The parameters in the ICMI model are obtained by fitting the known hadron spectra and they are taken directly from the previous work~\cite{Guo:2021mja}. The mass spectra, possible decay channels, and inner structures of the $S$-wave tetraquark $Qq\bar{Q}\bar{q}$ with quantum numbers $J^{PC} = 0^{+(+)}$, $1^{+(\pm)}$, and $2^{+(+)}$ are presented and analyzed. For the charge-neutral tetraquark state, the charge conjugation is also considered when analyzing the possible decay channels.

The recently observed hidden-charm tetraquark states, such as $Z_{cs}(3985)^-$, $X(3960)$, $Z_{cs}(4220)^+$, and so on, can be well explained in our model. The tetraquark state $cs\bar{c}\bar{u}$ with mass 3923.8 MeV and the quantum number $J^P = 1^+$ can be considered as a good candidate for $Z_{cs}(3985)^-$.
For the $cs\bar{c}\bar{s}$ configuration, we find a tetraquark state with a mass of 3976.5 MeV and a quantum number of $J^{PC}=0^{++}$. The properties of this state are in good agreement with $X(3960)$.
In the meantime, based on the wave functions of each $S$-wave tetraquark $Qq\bar{Q}\bar{q}$, we find the low-lying states in each configuration have a significant component of $|(Q_1\bar{Q}_3)^1(q_2\bar{q}_4)^1\rangle$ basis (the probability $|c_i|^2>90\%$) as shown in Table~\ref{atab4}. This indicates they have a large probability to decay into $Q\bar Q$ and $q\bar q$ mesons instead of $Q\bar q$ and $q\bar Q$ mesons. In some sense, these states are probably the molecule states of $Q\bar Q$ and $q\bar q$. Of course, other evidence, such as the mean radius and hadron level interactions, are needed to determine whether it is a molecular state~\cite{Voloshin:2006pz,Zhang:2006ix,Liu:2009qhy,Dong:2021lkh,Chen:2021cfl}. Our predictions on these exotic tetraquark states $Qq\bar{Q}\bar{q}$ can be examined in future experiments.

\begin{acknowledgments}
T. Guo is supported by the Scientific Research Foundation of Chengdu University of Technology under grant No.10912-KYQD2022-09557; J. Li and L. He are supported by the National Natural Science Foundation of China under grant No. 11890712; J. Zhao is supported by the European Union's Horizon 2020 research and innovation program under grant agreement No 824093 (STRONG-2020).
\end{acknowledgments}

%-------------------------------------------------------------------------------------------------------------------------------------
%\nocite{*}
%\bibliographystyle{unsrt}
%\bibliographystyle{plain}
\bibliography{QqQqref}

%\begin{thebibliography}{99}
%\end{thebibliography}
%--------------------------------------------------------------------------------------------------------------------------------

\end{document}